\documentclass[onecolumn,preprintnumbers,amsmath,prc,amssymb]{revtex4}

\usepackage{graphicx}
\usepackage{bm}

\begin{document}

\title{The critical indices of  the  Quark-Gluon Bags with Surface Tension Model
with  tricritical endpoint}

\author{A. I. Ivanytskyi}
\affiliation{Bogolyubov Institute for Theoretical Physics of the
National Academy of Sciences of Ukraine, Metrologichna str. 14$^b$, Kiev-03680, Ukraine
}

\begin{abstract}
The critical indices $\alpha'$, $\beta$, $\gamma'$ and $\delta$ of
the Quark Gluon Bags with Surface Tension  Model with the tricritical endpoint
are calculated as functions of the usual  parameters of this model  and two newly introduces parameters (indices). They are compared with the critical exponents  of  other models.
It is shown  that for  the newly introduced indices  $\chi = 0$ and $\xi^T \le 1$
there is a branch of  solutions for which
the critical exponents of the present model and the statistical multifragmentation model
coincide, otherwise  these models belong to different universality classes.
It is shown that for realistic values of the parameter $\varkappa$ the critical exponents
$\alpha'$, $\beta$, $\gamma'$ and $\delta$
of simple liquids and 3-dimensional Ising model can be only  described by  the branch
of solutions in which  all  indices except for $\alpha'$ correspond to  their values within the statistical multifragmentation model.
The scaling relations for the found critical exponents  are verified and  it is demonstrated  that for the standard definition of the index $\alpha'$ the Fisher and  Griffiths   scaling inequalities
 are not  fulfilled for some  values of  the model parameters, whereas the Liberman  scaling inequality is always obeyed.
Although it is shown that the specially defined index $\alpha'_s$ recovers the scaling
relations, another possibility, an existence of
 the non-Fisher universality classes,  is also discussed.

\vspace*{0.5cm}


\end{abstract}

\maketitle


\section{Introduction}

Investigation of the properties of strongly interacting matter equation of state has become a focal point of modern
nuclear physics of high energies. The low energy scan programs performed nowadays at CERN SPS and BNL RHIC
are aimed at the discovery of the (tri)critical endpoint of the quantum chromodynamics (QCD) phase diagram.
Despite many theoretical efforts neither an exact  location nor the properties of the QCD (tri)critical endpoint are well known \cite{Shuryak:sQGP}.  Therefore, the thorough theoretical  investigation of the QCD endpoint properties  are required in order  to clarify  whether this  endpoint is critical or  tricritical.

Here I would like to study the critical exponents of the QCD matter tricritical endpoint (triCEP).  Since such a  task, unfortunately, cannot be solved within the QCD itself and even the possibilities of the lattice QCD are nowadays  very limited in this respect, then,  unavoidably, one has to use some models of the triCEP.
The most popular models of this kind  are the quark-meson  model \cite{QMM:1,QMM:2}
and  the  extended Nambu--Jona-Lazinio model \cite{PNJL:1}. Their triCEPs are  generated by the
intersection of the deconfinement  phase transition (PT) line and the chiral symmetry restoration transition  line.
Since both kind of these models are the mean-field ones there is no reason to expect that their critical exponents
could  differ from that ones of the Van der Waals model  equation of state \cite{Stanley:71}.
Although very recently there appeared an interesting attempt to go beyond the mean-field
approximation in the quark-meson model \cite{QMM:3}, the resulting temperature of the triCEP is about 32 MeV only, which is too low and unrealistic compared to the lattice QCD predictions \cite{Misha:07}.
Therefore, to study the properties of non-classical triCEPs one has to investigate  the non-mean-field models.

Recently there appeared a comprehensive analysis \cite{CGreiner:10} of the QCD phase diagram with the triCEP within an extended version  of the gas of bags model \cite{GasOfBags:81},
which, however, is based on a physically  inadequate assumption.
Thus, in \cite{CGreiner:10} it is assumed
 that the Fisher topological exponent $\tau$ (denoted as $\alpha$ in
\cite{CGreiner:10}) of the Hagedorn mass spectrum  should depend on the baryonic chemical potential.
However,  there is no reason to believe that this exponent characterizing the fractal properties of the surface of  large bags should depend on chemical potential.
Moreover, the results on the size distribution of  clusters obtained for  the 2- and 3-dimensional  Ising model   by the Complement method
\cite{Complement} clearly show that even far away from the critical point  the Fisher  exponent $\tau$ has the same  value as at the endpoint.

Therefore, instead of model \cite{CGreiner:10} here   I consider another
extension of the gas of bags model \cite{GasOfBags:81} which is known as
the Quark Gluon Bags with Surface Tension Model  (QGBSTM)
\cite{Bugaev_07_physrev, Bugaev_07_physpartnucl}.
This exactly solvable model employes the same mechanism of the triCEP generation which is typical for the liquid-gas PT and which  is
used in the Fisher droplet model (FDM) \cite{Fisher_67} and in the statistical multifragmentation model (SMM) \cite{Bondorf,Bugaev_00}:
the endpoint of the 1-st order PT appears due to vanishing of the surface tension coefficient at this point which leads to the indistinguishability between the liquid and gas phases.
However,   the surface tension coefficient in the QGBSTM has the region of negative values, which is a principally different feature of this model  compared  to the FDM, SMM
and all other statistical  models of the liquid-gas PT.
Note  that just this feature  provides an existence of the cross-over at small values of baryonic chemical potential $\mu$ in the QGBSTM with triCEP  \cite{Bugaev_07_physrev}
and with CEP \cite{Bugaev_00}, and also it generates an  additional PT in the model with triCEP at large values of  $\mu$.
Therefore, it is very important and  interesting  to
study the critical indices of such a novel statistical model as  the QGBSTM,  to determine its class of universality and to examine  how the latter is related to  that  ones of the FDM and SMM.

The work is organized as follows. A brief description of the QGBSTM with the
triCEP  is given in Section \ref{secmodel}. In Section \ref{seccritical indices} the QGBSTM  is analyzed in details and its  critical exponents are calculated. Section \ref{secscallings} is devoted to the analysis of scaling relations between the found  critical exponents.
 Conclusions are given in Section \ref{secconclusions}.


%
\section{QUARK GLUON BAGS WITH SURFACE TENSION MODEL}
\label{secmodel}
An exact solution of the QGBSTM was found in \cite{Bugaev_07_physrev}.
The relevant degrees of freedom in this model are the quark gluon plasma (QGP) bags and hadrons.
The attraction between them is accounted like in the original statistical bootstrap model
\cite{SBM} via  many  sorts of the constituents, while the repulsion between them
is introduced a la Van der Waals equation of state
\cite{GasOfBags:81,Bugaev_07_physrev}.
An essential element of the QGBSTM  is  the $T$ and $\mu$ dependence of its  surface tension coefficient $T \Sigma(T_\Sigma,\mu)$ (here $\Sigma(T_\Sigma,\mu)$ is the reduced surface tension coefficient). Let us denote the nil line of the reduced surface tension
coefficient as  $T_\Sigma(\mu)$, i.e. $\Sigma(T_\Sigma,\mu)=0$.
Note that for a given $\mu$ the surface tension is negative (positive) for $T$ above (below) $T_\Sigma(\mu)$ line.

In the grand canonical ensemble the pressure of   QGP and hadronic phase are, respectively,  given by
\begin{eqnarray}
\label{pQ}
p_Q(T,\mu) & = & Ts_Q(T,\mu), \\
\label{pH}
p_H(T,\mu)& = & T\left[F_H(s_H,T,\mu)+u(T,\mu)I_\tau(\Delta s,\Sigma)\right],
\end{eqnarray}
where the following notations
\begin{eqnarray}\label{FH}
F_H(s_H,T,\mu) & = & \sum_{j=1}^n g_j e^{\frac{b_j \mu}{T} -v_js_H} \phi(T,m_j), \\
\label{FQ}
I_\tau(\Delta s,\Sigma) & = & \int\limits_{V_0}^{\infty}\frac{dv}{v^\tau}e^{-\Delta s v-\Sigma v^{\varkappa}},
\end{eqnarray}
are used. Here
$s_H\equiv\frac{p_H(T,\mu)}{T}$, $\Delta s\equiv s_H (T,\mu)-s_Q(T,\mu)$ and the particle density of a  hadron of mass $m_j$, baryonic charge $b_j$, eigenvolume $v_j$ and degeneracy $g_j$ is denoted as
$\phi_j(T,m_j)\equiv \frac{1}{2\pi^2}\int\limits_0^{\infty}\hspace*{-0.1cm}p^2dp\, e^{\textstyle-\frac{(p^2~+~m_j^2)^{1/2}}{T}}$.
The QGBSTM is solved  for a wide range of  the functions $u(T,\mu)$ and $s_Q(T,\mu)$  \cite{Bugaev_07_physrev}.
These functions are the parameters of the present model and here, as in \cite{Bugaev_07_physrev},
it is assumed  that the   functions  $u(T,\mu)$ and $s_Q(T,\mu)$ and   their first and second derivatives with respect to  $T$ and $\mu$   are finite  everywhere at the $T-\mu$ plane. As it was shown in \cite{Bugaev_07_physrev}, the 1-st  order deconfinement PT  exists for the  Fisher exponent $1<\tau \le  2$.
In the continuous part of the spectrum of bags
defined by (\ref{FQ}) the surface of a QGP bag of volume $v$ is parameterized by the term  $v^\varkappa$. Usually one chooses $\varkappa=\frac{2}{3}$ in 3-dimensional case (or $\varkappa=\frac{d-1}{d}$ for the  dimension $d$), but in what follows it  is  regarded as free parameter of  the range $0<\varkappa<1$.

\begin{figure}[ht]
\includegraphics[width=10cm, height=7cm]{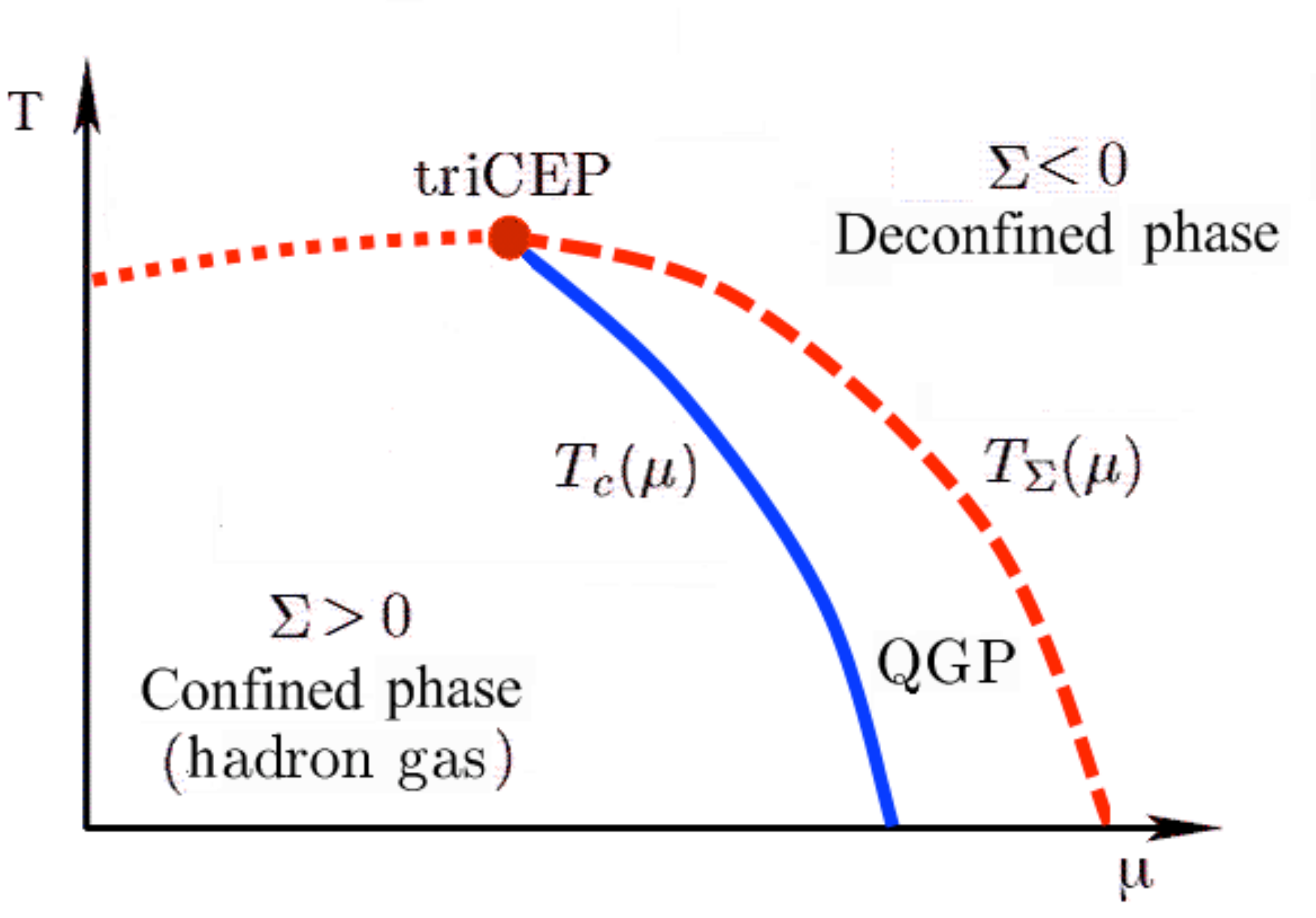}
\vspace*{-0.3cm}
\caption{[Color online]
A schematic  phase diagram in the plane of baryonic chemical potential $\mu$ and temperature $T$.
The dashed  curve indicates the nil line of the  surface tension coefficient  $T_\Sigma(\mu)$, below (above) which the surface tension  coefficient  is positive (negative).
The  deconfinement PT line  $T_c(\mu)$ is shown by the full curve  for $\mu > \mu_{cep}$
whereas the surface tension  induced  PT is depicted by  the long   dashed curve for $\mu > \mu_{cep}$. A cross-over (shown by the short dashed  curve) takes place along the line $T_\Sigma(\mu)$  for $\mu < \mu_{cep}$.  The  cross-over and PT regions are separated by triCEP (filled  circle).
At this  point  the curves $T_c(\mu)$ and $T_\Sigma(\mu)$ are tangent to each other.
}
\label{fig_T-Mu}
\end{figure}

The system pressure, that corresponds to a dominant  phase, is given by the largest value between $p_Q$ and $p_H$ for each set of $T$ and $\mu$. As usual,  the deconfinement PT occurs when pressure of  QGP  gets equal to that one of the hadron gas.
Suppose that the necessary conditions for the deconfinement PT outlined in
\cite{Bugaev_07_physrev} are satisfied and its  transition temperature is given by the function $T_c(\mu)$ for $\mu\ge\mu_{cep}$.
The necessary condition of the triCEP occurrence
is that at the nil line of the surface tension coefficient there exists the surface tension  induced
PT of 2-nd (or higher order)  \cite{Bugaev_07_physrev} for $\mu\ge\mu_{cep}$  and
 $T_c(\mu)\le T_\Sigma(\mu)$ for these $\mu$ values. The both of these  inequalities become equalities only at triCEP. Moreover,
at  triCEP
 the phase coexistence curve $T_c(\mu)$  is a tangent (not intersecting!) line to
 the nil line of the surface tension coefficient $T_\Sigma(\mu)$.
For  $\mu <  \mu_{cep}$ the deconfinement PT degenerates into a cross-over since in
this region $\Sigma(\mu)< 0$ \cite{Bugaev_07_physrev} and, hence, the system pressure
is defined  by a solution of  Eq.  (\ref{pH}).
A schematic picture of the phase diagram in $\mu-T$ plane is shown  in Fig. \ref{fig_T-Mu}.

An  actual parameterization of the reduced  surface tension coefficient $\Sigma(T,\mu)$ is taken from \cite{Bugaev_07_physrev} ($\zeta = const$):
\begin{equation}
\label{Sigma}
\Sigma(T,\mu)=\frac{\sigma_0}{T}\cdot\left| \frac{T_\Sigma(\mu)-T}{T_\Sigma(\mu)}\right|^\zeta  {\rm sign} \left(  T_\Sigma(\mu)-T \right)  \,,
\end{equation}
but here  it is written in a more general  form which allows one to perform
the calculations  for the integer and real  values of power $\zeta$. In what follows the
 coefficient $\sigma_0$ is assumed to be a positive constant, i.e.  $\sigma_0={\rm const}>0$,  but  it is easy to show that   the obtained results hold, if  $\sigma_0 >0$ is  a smooth function of $T$ and $\mu$.  Here it is appropriate to say a few words about the negative values of $\Sigma$.  Note that the existence of regions with  $\Sigma < 0$ is a distinctive feature of QGBSTM compared to other models. There is nothing wrong or unphysical with the negative values of surface tension coefficient, since $T \Sigma\, v^\varkappa$ is the surface  free energy of the
bag of mean volume $v$ and, hence, as any free energy,  it contains the energy part
$e_{surf}$ and  the entropy part $s_{surf}$ multiplied by temperature $T$ \cite{Fisher_67}. Therefore, at low temperatures the energy part dominates and surface  free energy is positive, whereas at high temperatures the number of bag configurations with large surface  drastically increases  and it exceeds  the  Boltzmann suppression and, hence,
  the surface free energy becomes negative since $s_{surf} > \frac{e_{surf}}{T}$.
Such a behavior of the surface free energy  can be derived within the exactly solvable model of surface deformations
known as Hills and Dales Model \cite{Bugaev_05_Physrev}.  Moreover, very recently
using
the relation between  the tension of  confining color string and the surface tension of the QGP bag  derived in \cite{String:09} it was possible to demonstrate
\cite{String:09,String:11a} that at the cross-over region the surface tension coefficient of large bags  is unavoidably negative. In addition this approach allows  one to determine
such an important parameter of the QGBSTM as the value of   triCEP/CEP   temperature
$T_{cep} = 153.9 \pm 4.5 $ MeV  \cite{String:11b} under the plausible
assumptions which are typical for the liquid-gas PT   \cite{Fisher_67,Bugaev_05_Physrev}.
Note that this value of the  triCEP/CEP    temperature is in a good agreement with the
lattice QCD results discussed in \cite{Misha:07}.


In the vicinity of triCEP
the behavior  of both  the deconfinement PT curve and the nil  surface tension coefficient line in the $\mu-T$ plane is  parameterized via a single  parameter $\xi^T>0$:
\begin{equation}
\label{xiT}
T_{cep}-T_\Sigma(\mu)\sim T_{cep}-T_c(\mu)\sim(\mu-\mu_{cep})^{\xi^T} \,,
\end{equation}
since, as discussed above,   $T_c(\mu)$ and $T_\Sigma(\mu)$ are tangent to each other  at triCEP. This is a new index which was not considered both in the FDM and SMM.
As it will be shown below this index  is responsible for a new universality class compared to other
exactly solvable models.

At the phase coexistence curve
the Clapeyron-Clausius equation can be written as  $\frac{d\mu_c}{dT}=-\frac{S_H-S_Q}{\rho_H-\rho_Q}\Bigl|_{T=T_c}$.  The entropy density  $S_A$ and the  baryonic density $\rho_A$ are, respectively,  defined as $T$ and $\mu$ partial derivatives of
the corresponding pressure $p_A$,  where $A \in \{H, Q\}$. Then using  (\ref{pQ}) and (\ref{pH}) the Clapeyron-Clausius equation can be explicitly  rewritten as
\begin{equation}\label{EqVII}
\frac{d\mu_c}{dT}=-\frac
{\frac{\partial F_H}{\partial T}+\frac{\partial u}{\partial T}I_\tau(0,\Sigma)+\frac{\partial s_Q}{\partial T}
\left(\frac{\partial F_H}{\partial s}-1\right)-\frac{\partial\Sigma}{\partial T}uI_{\tau-\varkappa}(0,\Sigma)}
{\frac{\partial F_H}{\partial\mu}+\frac{\partial u}{\partial\mu}I_\tau(0,\Sigma)+\frac{\partial s_Q}{\partial\mu}
\left(\frac{\partial F_H}{\partial s}-1\right)-\frac{\partial\Sigma}{\partial\mu}uI_{\tau-\varkappa}(0,\Sigma)}
\Biggl|_{T=T_c}.
\end{equation}
Let us   parameterize    the  behavior of  the numerator and denominator in
(\ref{EqVII}) at  the triCEP vicinity as
\begin{eqnarray}
\left[\frac{\partial F_H}{\partial T}+\frac{\partial u}{\partial T}I_\tau(0,\Sigma)+\frac{\partial s_Q}{\partial T}
\left(\frac{\partial F_H}{\partial s}-1\right)\right]_{T=T_c} & \sim & (T_{cep}-T_c(\mu))^{\chi+\frac{1}{\xi^T}-1},
\\
\label{combinationmu}
\left[\frac{\partial F_H}{\partial\mu}+\frac{\partial\,  u}{\partial\mu}I_\tau(0,\Sigma)+\frac{\partial s_Q}{\partial\mu}
\left(\frac{\partial F_H}{\partial s}-1\right)\right]_{T=T_c}& \sim & (T_{cep}-T_c(\mu) )^\chi,
\end{eqnarray}
where  $\chi\ge\max(0,1-\frac{1}{\xi^T})$ denotes  another  new index. The latter inequality  follows from the fact that the integral $I_\tau$ and functions $F_H$, $u$, $s_Q$ together  with their derivatives are finite for any finite values of  $T$ and $\mu$.
It is necessary  to emphasize that the index  $\chi$  unavoidably  appears  from an inspection of the  Clapeyron-Clausius equation.  Since the latter  is a direct  consequence of the  Gibbs criterion for  phase equilibrium, then an introduction of the parameter  $\chi$   is quite general and could be done for any model with the  PT of the liquid-gas type.
So far,  the index   $\chi$ was never used for the calculation of critical exponents.
For example,  in the  SMM case   this parameter is  set  to  zero by construction \cite{Reuter,Bugaev:2010xm}.


%
\section{THE CRITICAL INDICES OF THE QGBSTM}
\label{seccritical indices}

The standard set of  critical exponents  $\alpha'$, $\beta$ and $\gamma$
\cite{Fisher_70,Stanley:71} describes the $T$-dependence of the system near triCEP:
\begin{eqnarray}
\label{alphadef}
C_\rho  & \sim\ &
|t|^{-\alpha'},  \hspace*{0.1cm}
{\rm for}  \quad  t \le 0\quad{\rm and}\quad\rho=\rho_{cep},\\
\Delta\rho & \sim  & |t|^\beta,  \hspace*{0.4cm}
{\rm for}  \quad t \le 0, \\
\Delta K_T & \sim  & |t|^{-\gamma'},  \hspace*{0.1cm}
{\rm for}  \quad t < 0,
\end{eqnarray}
where $\Delta\rho\equiv(\rho_Q-\rho_H)_{T=T_c}$ defines the order parameter, $C_\rho\equiv\frac{T}{\rho}(\frac{\partial S}{\partial T})_\rho$ denotes the specific heat at the critical density and $\Delta K_T\equiv(K_T^H-K_T^Q)_{T=T_c}$ is the discontinuity in the isothermal compressibility $K_T\equiv\frac{1}{\rho}(\frac{\partial\rho}{\partial p})_T$ across the PT line, the variable $t$ is the  reduced temperature $t \equiv \frac{T-T_{cep}}{T_{cep}}$.
The   critical isotherm shape is given by the  index $\delta$ \cite{Fisher_70,Stanley:71}
\begin{equation}
p_{cep}-\widetilde{p}\sim(\rho_{cep}-\widetilde{\rho})^\delta  \hspace*{0.1cm}
{\rm for}  \quad t = 0.
\end{equation}
Hereafter the tilde indicates that $T=T_{cep}$.

%
%


The calculation of $\alpha'$ requires the  knowledge of  the specific heat behavior $C_\rho\equiv\frac{T}{\rho}(\frac{\partial S}{\partial T})_\rho$ along the critical isochore  $\rho=\rho_{cep}$ inside the mixed  quark-gluon-hadron phase.
In contrast to the works \cite{Reuter,Fisher_70},
  the famous Yang-Yang formula \cite{YangYang:64} is  not used here to calculate $C_\rho$, since this formula   leads to a less convenient representation.  Therefore, below there are given some details of the specific heat evaluation.

As usual
the
entropy density  and  baryonic density of the mixed phase are defined via that ones  of the pure phases and the parameter $\lambda$, which is the  volume fraction of hadronic phase (i.e. the volume fraction of QGP is, respectively, $1-\lambda$):
\begin{eqnarray}
\label{rhoMixedPhase}
\rho|_{T=T_c} & = & \lambda\rho_H|_{T=T_c}+(1-\lambda)\rho_Q|_{T=T_c}, \\
\label{SMixedPhase}
S|_{T=T_c} & = & \lambda S_H|_{T=T_c}+(1-\lambda)S_Q|_{T=T_c}.
\end{eqnarray}
Varying  $\lambda$ from 0 to 1 one can describe any state in  the mixed phase for fixed $\mu $ or $T$, since these variables  are not independent inside the mixed phase. Replacing   $\rho|_{T=T_c}$ by $\rho_{cep}$  in Eq. (\ref{rhoMixedPhase}),  one can  calculate the  total $T$-derivative of $\lambda$ along the critical isochore $\rho=\rho_{cep}$. Then for $\rho=\rho_{cep}$  from  Eq. (\ref{SMixedPhase}) one finds
\begin{equation}\label{EqXVI}
C_{\rho}=\frac{T}{\rho_{cep}}
\left[\lambda\left(\frac{d}{dT}(S_H-S_Q)_{T=T_c}+\frac{d\mu_c}{dT}\frac{d}{dT}(\rho_H-\rho_Q)_{T=T_c}\right)+
\left(\frac{d(S_Q|_{T=T_c})}{dT}+\frac{d\mu_c}{dT}\frac{d(\rho_Q|_{T=T_c})}{dT}\right)\right].
\end{equation}
Using  the Clapeyron-Clausius equation
it is possible to rewrite the specific heat  (\ref{EqXVI})  in the form
\begin{equation}
\label{Crho}
C_{\rho}=\frac{T}{\rho_{cep}}
\left[(\rho_Q-\rho_{cep})_{T=T_c}\frac{d^2\mu_c}{dT^2}+
\left(\frac{d(S_Q|_{T=T_c})}{dT}+\frac{d\mu_c}{dT}\frac{d(\rho_Q|_{T=T_c})}{dT}\right)\right],
\end{equation}
where the definition  (\ref{rhoMixedPhase}) for  $\lambda$ was used.

The critical exponent $\alpha'$ describes the  temperature behavior of the most singular term in Eq. (\ref{Crho}).
Expanding  $\rho_Q|_{T=T_c}$ into series of $t$-powers  and using the  parametrization
(\ref{xiT})  of the coexistence curve one can show  that the  first term in  (\ref{Crho}) behaves as $t^{\min(1,\frac{1}{\xi^T})+\frac{1}{\xi^T}-2}$.
Since the
entropy density and baryonic  density of QGP are defined via the $T$ and $\mu$ derivatives of the function $Ts_Q(T,\mu)$, which is a  regular function together   with its first and second  derivatives,   then  the singularity in  the second term of  (\ref{Crho})
appears from $\left(\frac{d\mu_c}{dT}\right)^2 \sim t^{\frac{2}{\xi^T}-2}$. Clearly,
the  second term  in (\ref{Crho})  is  a  non-vanishing constant,  if  $T$-derivative of $\mu_c(T)$ is finite at triCEP. Accounting for this fact,  one gets   the  critical exponent $\alpha'$  as
\begin{equation}
\alpha'=2-2\min\left(1,\frac{1}{\xi^T}\right).
\end{equation}
This equation shows
 that $\alpha'>0$ for $\xi^T>1$ only,  otherwise  $ \alpha'=0$.
As it will be seen below only  the index $\alpha' > 0$ depends on $\xi^T$,
whereas other critical exponents always have the branch of solutions  which is independent on $\xi^T$.  Compared  to the FDM and SMM, such a property leads to a principally new
and unique possibility  to choose the  index $\alpha'$  independently of other critical indices.


To calculate the critical exponents $\beta$ and $\gamma'$
it is necessary to analyze the
behavior of the integral $I_{\tau- q}(0,\Sigma)$   for small positive values of $\Sigma$,  which is the case,   when $T=T_c(\mu)$ approaches $T_{cep}$.
In the limit $\Sigma\rightarrow+0$ the  integral $I_{\tau- q}(0,\Sigma)$ remains finite for $\tau>1+q$ and diverges otherwise. For $\tau=1+q$ it  diverges  logarithmically.
Indeed,
the substitution $z\equiv v \, \Sigma^{\frac{1}{\varkappa}}$ yields
\cite{Reuter, Bugaev:2010xm}
\begin{equation}\label{EqXIX}
I_{\tau-q}(0,\Sigma)=\Sigma^{\frac{\tau-1- q}{\varkappa}}\int\limits_{V_0\Sigma^{\frac{1}{\varkappa}}}^{\infty}
dz~\frac{e^{-z^\varkappa} }{z^{\tau-q}}~.
\end{equation}
%


The condition $\tau<q+1$ guaranties a convergence of the integral (\ref{EqXIX}) at its lower limit. The finite values of $I_{\tau-q}(0,\Sigma)$ at the upper limit is provided  by  an exponential factor. Summarizing  these  results  for $\Sigma\rightarrow+0$ one can write
\begin{eqnarray}
\label{I(tau-q)(0,S)}
I_{\tau- q}(0,\Sigma)\sim\
\left\{ \begin{array}{lr}
\Sigma^{\frac{\tau-1- q}{\varkappa}},  &\hspace*{0.1cm}
{\rm for} \quad \tau<1+q\\
\ln\Sigma,  &\hspace*{0.1cm}
{\rm for}  \quad  \tau=1+q\\
{\rm const},  &\hspace*{0.1cm}
{\rm for}  \quad  \tau>1+q\\
\end{array}\right.~\sim~\Sigma^{\min\left(0,\frac{\tau-1- q}{\varkappa}\right)},
\end{eqnarray}
where the limit  $(x^y\ln x)_{x\rightarrow0}\sim x^y$ was used.

To  calculate the index $\delta$ one should  investigate the integral
(\ref{FQ}) in the limit  $\widetilde{\Delta} s\rightarrow+0$ and $\widetilde{\Sigma} \rightarrow -0$, but for a condition  $\frac{\widetilde{\Sigma}}{\widetilde{\Delta} s^\varkappa} \rightarrow -0$ which, as shown in \cite{Bugaev_07_physrev}, is the case  (also see  Eq. (\ref{EqXXVIII}) for details). Then one obtains
\begin{eqnarray}
\label{I(tau-q)(D,S)}
I_{\tau- q}(\widetilde{\Delta} s,\widetilde{\Sigma})\sim\
\left\{ \begin{array}{lr}
\widetilde{\Delta} s^{\tau-1- q},  &\hspace*{0.1cm}
{\rm for} \quad \tau<1+q\\
\ln\widetilde{\Delta} s,  &\hspace*{0.1cm}
{\rm for}  \quad  \tau=1+q\\
{\rm const},  &\hspace*{0.1cm}
{\rm for}  \quad  \tau>1+q\\
\end{array}\right.~\sim~\widetilde{\Delta} s^{\min\left(0,\tau-1- q\right)}.
\end{eqnarray}
%


Using the definition of the baryonic density and  Eqs. (\ref{pQ}) and  (\ref{pH}) one can find the baryonic density discontinuity across the deconfinement PT line:
\begin{equation}
\label{deltarho}
\Delta\rho=T_c\cdot\frac
{\frac{\partial F_H}{\partial\mu}+\frac{\partial u}{\partial\mu}I_\tau(0,\Sigma)+\frac{\partial s_Q}{\partial\mu}
\left(\frac{\partial F_H}{\partial s}-1\right)-\frac{\partial\Sigma}{\partial\mu}uI_{\tau-\varkappa}(0,\Sigma)}
{1-\frac{\partial F_H}{\partial s}+uI_{\tau-1}(0,\Sigma)}\biggl|_{T=T_c} \, .
\end{equation}
According to  the  parameterizations (\ref{Sigma})  and  (\ref{xiT}) one obtains  that along 
the PT line $\Sigma\sim t^{\zeta}$ and
$\frac{\partial \Sigma}{\partial\mu}\sim t^{\zeta-\frac{1}{\xi^T}}$ and, hence,  the temperature dependence  of
$\Delta\rho$ in (\ref{deltarho}) can be straightforwardly  found  from Eqs. (\ref{combinationmu}) and (\ref{I(tau-q)(0,S)}). This yields
\begin{eqnarray}
\label{beta}
\beta=\frac{\zeta}{\varkappa}(2-\tau)+
\left\{ \begin{array}{lr}
\chi,    &\hspace*{0.1cm}
{\rm for} \quad \chi\le\frac{\zeta}{\varkappa}\min(\varkappa,\tau-1)-\frac{1}{\xi^T}\\
\frac{\zeta}{\varkappa}\min(\varkappa,\tau-1)-\frac{1}{\xi^T},    &\hspace*{0.1cm}
{\rm for} \quad \chi\ge\frac{\zeta}{\varkappa}\min(\varkappa,\tau-1)-\frac{1}{\xi^T}
\end{array}\right..
\end{eqnarray}
The structure of this equation is  obvious: its first term corresponds to the denominator of Eq. (\ref{deltarho}), whereas the  second one describes the leading  term of the  numerator of   (\ref{deltarho}).


To find  the
critical exponent $\gamma'$ one has to calculate the  isothermal compressibility $K_T\equiv\frac{1}{\rho}(\frac{\partial\rho}{\partial p})_T$ both  for  QGP and for hadronic phase.
With the help of the baryonic density definition one can rewrite the isothermal compressibility as  $K_T=\frac{1}{\rho^2}\frac{\partial^2p}{\partial\mu^2}$. This  result  clearly  demonstrates that the QGP contribution into the $\Delta K_T$ is negligibly small since the QGP pressure   is given by the function $s_Q(T,\mu)$, which,
according to the  QGMSTM assumptions \cite{Bugaev_07_physrev},   is finite together with its first and second derivatives for all finite values of $T$ and $\mu$.  Therefore, at triCEP $\Delta K_T$ can  diverge  due to the hadronic phase contribution only, i.e.  near triCEP $\Delta K_T\simeq K_T^H$. Therefore, using Eq. (\ref{pH}) and keeping at triCEP  the  most
singular  terms one finds
\begin{equation}
\Delta K_T\simeq\left[\frac{T}{\rho_H^2}\frac
{\left(\frac{\partial\Delta s}{\partial\mu}\right)^2uI_{\tau-2}(0,\Sigma)}
{1-\frac{\partial F_H}{\partial s}+uI_{\tau-1}(0,\Sigma)}
\right]_{T=T_c(\mu)}\sim~
\frac{I_{\tau-2}(0,\Sigma)}{I_{\tau-1}(0,\Sigma)}\cdot\left(\frac{\partial\Delta s}{\partial\mu}\right)_{T=T_c(\mu)}^2.
\end{equation}
From the definition of $\Delta s$  it follows that along the PT line $\frac{\partial\Delta s}{\partial\mu}=\frac{\rho_H-\rho_Q}{T}\sim t^\beta$. Then from  Eqs. (\ref{I(tau-q)(0,S)}), (\ref{Sigma}), (\ref{xiT}) one has
\begin{equation}\label{gamma}
\gamma'=\frac{\zeta}{\varkappa}-2\beta.
\end{equation}
%

At  the  critical isotherm one can use  the definition of $\Delta s$ to get
\begin{eqnarray}
\widetilde{p}-p_{cep}=T_{cep}(\widetilde{\Delta} s+\widetilde{s}_Q-s_Q|_{cep})=
T_{cep}\left(\widetilde{\Delta} s+\Delta\mu\frac{\partial s_Q}{\partial\mu}\biggl|_{cep}\right),
\end{eqnarray}
where in the second step one has to expand  $\widetilde{s}_Q$  in powers of $\Delta\mu \equiv \mu - \mu_{cep}$ and keep the linear  term.
Similarly one determines the deviation of the
baryonic density   taken at the critical isotherm that is  lying  outside the mixed phase
from  the baryonic density  at triCEP:
\begin{equation}
\widetilde{\rho}-\rho_{cep}=T_{cep}\left(\frac{\partial\widetilde{\Delta} s}{\partial\mu}+
\frac{\partial\widetilde{s}_Q}{\partial\mu}-\frac{\partial s_Q}{\partial\mu}\biggl|_{cep}\right)=
T_{cep}\left(\frac{\partial\widetilde{\Delta} s}{\partial\mu}+
\Delta\mu\frac{\partial^2s_Q}{\partial\mu^2}\biggl|_{cep}\right).
\end{equation}
Now it is clear that the behavior of $\widetilde{\Delta} s$ near triCEP should be analyzed in
order
to calculate the critical exponent $\delta$.
 Let us consider the case $\chi=0$ first. Such a case  is typical for the  SMM \cite{Bugaev_00,Reuter}.
 Since  at triCEP $\widetilde{\Delta} s=0$, then substituting  the expansion $I_\tau(\widetilde{\Delta} s,\widetilde{\Sigma}) \approx I_\tau(0,0)-
\widetilde{\Delta} sI_{\tau-1}(\frac{\widetilde{\Delta} s}{2},\frac{\widetilde{\Sigma}}{2})-
\widetilde{\Sigma}I_{\tau-\varkappa}(\frac{\widetilde{\Delta} s}{2},\frac{\widetilde{\Sigma}}{2})$ \cite{Bugaev_07_physrev} into Eq. (\ref{pH}) for the  cross-over states one gets
\begin{equation}\label{EqXXVIII}
\widetilde{\Delta} s=\frac
{\Delta\mu\left(\frac{\partial F_H}{\partial\mu}+\frac{\partial u}{\partial\mu}I_\tau(0,0)+
\frac{\partial s_Q}{\partial\mu}\left(\frac{\partial F_H}{\partial s}-1\right)\right)_{cep}
-~\widetilde{\Sigma}\, \widetilde{u}\, I_{\tau-\varkappa}(\frac{\widetilde{\Delta} s}{2},\frac{\widetilde{\Sigma}}{2})}
{1-\frac{\partial\widetilde{F}_H}{\partial\mu}+
\widetilde{u}\,I_{\tau-1}(\frac{\widetilde{\Delta} s}{2},\frac{\widetilde{\Sigma}}{2})} \,,
\end{equation}
where it is sufficient to  keep only the first order terms of expansion, whereas for the case $\chi>0$ one has to keep  the  second order terms as well. With the help of  Eq. (\ref{I(tau-q)(D,S)}) one concludes that
\begin{eqnarray}
\widetilde{\Delta} s\sim\
\left\{ \begin{array}{lr}
\Delta\mu^{\frac{\xi^T\,\zeta}{\max(\tau-1,\varkappa)}},  &\hspace*{0.1cm}
{\rm for} \quad \frac{\xi^T\,\zeta}{\max(\tau-1,\varkappa)} \le \frac{1}{\tau-1}\\
\Delta\mu^{\frac{1}{\tau-1}},  &\hspace*{0.1cm}
{\rm for}  \quad  \frac{\xi^T\,\zeta}{\max(\tau-1,\varkappa)} \ge \frac{1}{\tau-1}\\
\end{array}\right.,
\end{eqnarray}
which allows one to find the index  $\delta$ for $\chi=0$
\begin{eqnarray}\label{EqXXX}
\delta |_{\chi=0} =
\left\{ \begin{array}{lr}
\left[\frac{\xi^T\,\zeta}{\max(\tau-1,\varkappa)}-1\right]^{-1},  &\hspace*{0.1cm}
{\rm for} \quad 1<\frac{\xi^T\,\zeta}{\max(\tau-1,\varkappa)} \le \frac{1}{\tau-1}\\
\frac{\tau-1}{2-\tau},  &\hspace*{0.1cm}
{\rm for}  \quad  \frac{\xi^T\,\zeta}{\max(\tau-1,\varkappa)} \ge \frac{1}{\tau-1}\\
\end{array}\right..
\end{eqnarray}
Note that in this case   the  inequalities  $\frac{3}{2}<\tau \le 2$ providing  the existence of the 2-nd  order PT at triCEP \cite{Bugaev_07_physrev} also  guaranty the fulfillment of the condition  $  \delta>1$.

The result for the index  $\delta$ in the  case $\chi>0$   can be found similarly:
\begin{equation}\label{Eqdelta}
\delta |_{\chi>0} =\left[\frac{\xi^T\,\zeta}{\max(\tau-1,\varkappa)}-1\right]^{-1}\,, \quad {\rm for} {\textstyle  \quad1<\frac{\xi^T\,\zeta}{\max(\tau-1,\varkappa)} \le \frac{2}{\tau-1} } \, .
\end{equation}

Since in some aspects the QGBSTM is similar to the FDM and SMM it
is interesting to compare its critical exponents  with that ones of the FDM \cite{Fisher_67} and  SMM \cite{Reuter}. However, one has also   to remember that
in contrast  to the FDM, both the  QGBSTM and the SMM have a non-vanishing excluded volume of the constituents and the same range  of  the Fisher exponent $1<\tau \le 2$ for
the triCEP existence, whereas  the  FDM  can be formulated for  $\tau>2$ only.
Since  the FDM and SMM implicitly treat the parameter $\chi=0$,  then it is most  natural
to compare
 their  critical exponents with the QGBSTM results   just  for such a  case.
Eq. (\ref{EqXXX}) shows that in the QGBSTM there is a regime, when its
index  $\delta |_{\chi=0}$ matches the SMM result \cite{Reuter}.
Moreover,   it is easy to see that in this regime  all other indices
 the QGBSTM and SMM coincide for $\xi^T \le 1$
\begin{eqnarray}\label{EqXXXII}
\alpha' = 0 \,,  \quad \beta =  \frac{\zeta}{\varkappa} (2 - \tau)\,, \quad   \gamma' =  \frac{2 \, \zeta}{\varkappa}
\left( \tau - \frac{3}{2}  \right)\, ~ ~{\rm and} \quad  \delta |_{\chi=0}=\frac{\tau-1}{2-\tau} \,,
\end{eqnarray}
although the present model has  entirely  new regime for $\xi^T > 1$, for which  all  its indices except for $\alpha'$ are equal to the corresponding exponents  of the SMM,  whereas
$\alpha'$ can be chosen freely provided that   the index  $\xi^T$ exceeds the value $ \frac{\max(\tau-1,\varkappa)}{\zeta\, (\tau-1)} $.
On the other hand the QGBSTM  always belongs to a different universality
class $1 < \tau \le 2$ (except for a special case $\tau=2$),  than that one of  FDM in which $\tau \ge  2$.
Thus, the spectrum of values of  the  QGBSTM critical indices is  more rich than the corresponding spectra   of the FDM and SMM since this model contains two new indices
$\xi^T$ and $\chi$.

Let us demonstrate this using a few  sets of critical indices. Since the present model contains five parameters, $\tau$,
$\varkappa$, $\zeta$,  $\xi^T$ and $\chi$, there exists an infinite number of possibilities to describe the four standard critical exponents,  $\alpha'$, $\beta$, $\gamma'$ and
$\delta$. Thus, the critical exponents of the 2-dimensional Ising model \cite{Huang} shown in the first row of the Table I can be exactly  reproduced by many sets of parameters with the vanishing value of  index $\chi$ (see the first row in the Table II).
It is interesting to mention that, if one uses the linear $T$-dependence of the surface tension coefficient (\ref{Sigma}) employed in the FDM for $T \le T_{cep}$, i.e. fixes  $\zeta=1$, then for
 the 2-dimensional Ising model one obtains $\varkappa = \frac{1}{2}$, which is typical for  the dimension $d=2$. However, in contrast to the FDM,  where $\tau = \frac{31}{15}>2$, this set of critical exponents is described by the value  $\tau = \frac{31}{16}<2$.

It is interesting to analyze the critical indices of the simple liquids since near the deconfinement region
the QGP  behaves as  a strongly interacting liquid \cite{Shuryak:sQGP}. Taking the wide range of values for the critical exponents of simple liquids \cite{Huang} (see the Table I) one can describe them in many ways.
The set A  in the Table II demonstrates one of  such  possibilities, but its value of the parameter $\varkappa = (\tau-1)
\frac{437}{720} \le \frac{437}{720} \approx 0.6069$  is essentially smaller than the expected for 3-dimensions  value  $\varkappa \approx  \frac{2}{3}$.  At the first glance it seems that the accuracy of above 10\% for the set A value of  $\varkappa$  which is obtained   for $\tau < 2$ is acceptable, but a careful  study of the surface free energy in various cluster models \cite{Fisher_67,Fisher_70,Elliott:06} shows that for 3-dimensional case
the value $\varkappa = \frac{2}{3}$ holds with much better accuracy of about 4\%.
However, if one requires that  the critical exponents of simple liquids are reproduced for  $\varkappa >  0.49$, i.e. in such a way that both 2- and 3-dimensional  values of parameter  $\varkappa$ can be included, then
it can be shown that the only existing  solution corresponds to the SMM values for
indices $\beta$, $\gamma'$ and $\delta$ (\ref{EqXXXII}) with $\chi=0$,
while the index $\alpha'$  is fixed to its experimental value (see the set B for liquids  in the Table II).
In fact,  the same outcome  is obtained   for  the critical exponents of the 3-dimensional Ising model (see the Table II) which were found with very high accuracy in \cite{Campostrini:05} and are given in the Table I.

There are two important consequences that follow from these results. First,  since  the QGBSTM indices $\beta$,  $\gamma'$ and $\delta |_{\chi=0}$ match  that  ones of the SMM and the latter do not depend on the parameter $\xi^T$, then  there exist two relations for index $\tau$ \cite{Reuter}
\begin{equation}\label{EqXXXIII}
\tau = 2 - \frac{1}{1+\delta}~~~{\rm and}~~~\tau = 2 - \frac{\beta}{\gamma'+2 \beta}\, .
\end{equation}
The first of these equalities follows from the definition of index  $\delta |_{\chi=0}=\frac{\tau-1}{2-\tau}$, whereas the second one can be obtained directly from Eq.
(\ref{EqXXXII}).  Second,
from  (\ref{EqXXXIII}) it follows that the  index $\tau \approx 1.826 \pm 0.02$ has a very narrow range  for the set B of  simple liquids   and for
the 3-dimensionl Ising model (see the Table II). Both of these consequences are important for  QCD since
its universality  class  is expected to match the class  of the 3-dimensionl Ising model \cite{Misha:07,  RandomMatr:98, Rajagopal:99}. Therefore, on the basis of above results one can expect \cite{String:11b} that at triCEP the volume  distribution of large  QGP bags  has a power like form
$v^{-\tau}$ with $\tau \approx 1.826 \pm 0.02$.
The same conclusion for the mass distribution results from the fact that in the QGBSTM \cite{Bugaev_07_physrev}
the mass of large  bags and its volume $v > V_0$  are proportional to each other.

\begin{table}[t]
\bigskip
\begin{tabular}{|c|c|c|c|c|c|c|c|c|c|c|c|c|c|}
\hline
        \hspace*{1.0cm}    &        $\alpha'$       &     $\beta$     &   $\gamma'$   &  $\delta$   \\ \hline

      2d Ising model       &            0           &  $\frac{1}{8}$  & $\frac{7}{4}$ &     15      \\  \hline
      Simple liquids        &        0.09-0.11       &    0.32-0.35    &    1.2-1.3    &  4.2-4.8    \\  \hline
      3d Ising model       & $0.1096\pm0.0005$ & $0.3265\pm0.0001$ & $1.2373\pm0.0002$ & $4.7893\pm0.0008$ \\  \hline
\end{tabular}
\vspace*{0.3cm}
\caption{
The critical indices of simple liquids \cite{Huang}, 2-dimensional Ising model \cite{Huang} and 3-dimensional Ising model \cite{Campostrini:05}.
}
\label{table1}
\end{table}


%
\begin{table}[t]
\bigskip
\begin{tabular}{|c|c|c|c|c|c|}
\hline
        \hspace*{1.0cm}    &        $\chi$       &     $\xi^T$     &   $\tau$   &  $\varkappa$  &  $\zeta$ \\ \hline
      2d Ising model       &            0           &  $\frac{8}{15} \le\xi^T \le 1  $  & $\frac{31}{16}$ &    $\min(2\varkappa,\frac{15}{8})\ge\frac{1}{\xi^T}$   & $2 \varkappa $    \\  \hline
      Simple liquids  A     &       $0 <  \chi  <  \frac{8}{23}  $      &    $\frac{20}{19}$    &    $\tau=\frac{20}{11}+\chi\frac{23}{44}$   & $(\tau-1)\frac{437}{720}$ &  $\frac{44}{23} \varkappa$ \\  \hline
     Simple liquids  B     &        0    & $1.0526\pm0.0055$  & $1.8255\pm0.0212$  & $\varkappa\ge0.4947$ & $\varkappa\cdot(1.92\pm0.026)$
   \\  \hline
      3d Ising model       &     0    & $1.0579\pm0.000055$ & $1.8272\pm0.000048$ & $\varkappa\ge0.4999$ &  $\varkappa\cdot(1.8903\pm0.00007)$          \\  \hline
\end{tabular}
\vspace*{0.3cm}
\caption{
The QGBSTM parameters that describes the corresponding exponents given in the Table I.
}
\label{table2}
\end{table}


%
\section{The Scaling relations of the QGBSTM}
\label{secscallings}

The well known exponent inequalities were   proven for real gases by
\begin{eqnarray}
\label{Fisher}
{\rm Fisher}\, [26]:& \quad \alpha'+2\beta+\gamma'\ge 2, \\
\label{Griffiths}
{\rm Griffiths}\, [27]:& \quad \alpha'+\beta(1+\delta)\ge 2,\\
\label{Liberman}
{\rm Liberman}\, [28]:& \quad \gamma'+\beta(1-\delta)\ge 0.
\end{eqnarray}
The corresponding exponent inequalities for magnetic systems are often called   Rushbrooke, Griffiths and  Widom  inequalities, respectively.
These inequalities are traditionally believed to play a fundamental role in the modern theory of  critical phenomena. However, the real situation with the scaling inequalities (\ref{Fisher})--(\ref{Liberman})  is not
that trivial as it is often presented in the textbooks.
Thus, as one can see from the Table III the inequalities  (\ref{Fisher})--(\ref{Liberman}) exactly hold for the Onsager solution only, whereas for simple liquids only the Fisher relation is obeyed
and even for highly accurate numerical evaluation of the critical exponents of the 3-dimensional Ising model there are some problems with  the inequalities (\ref{Fisher}) and
(\ref{Liberman}).
Moreover, a  long time ago it was found
\cite{Fisher_70} that
the traditional  definition of the exponent $\alpha'$ given by (\ref{alphadef})  may lead to somewhat smaller value than 2 staying   on  the right hand side of Eqs. (\ref{Fisher}) and (\ref{Griffiths}).   A similar  result was analytically  found for the SMM \cite{Reuter, Bugaev:2010xm}, which shows that for the standard set of the SMM parameters \cite{Bondorf, Bugaev_00} the right hand side of inequalities (\ref{Fisher})  and (\ref{Griffiths}) should be replaced by $\frac{15}{8}$. Therefore, it is interesting to verify the scaling inequalities for the QGBSTM  indices obtained here.


Despite the usual  expectations,
the  QGBSTM critical exponents  do  not obey  the traditional scaling relations  in general. Again, as in  the SMM case,  the  Fisher  and Griffiths  inequalities
are not always  fulfilled, whereas
the Liberman  inequality is fulfilled for any values of the model parameters.
Indeed, let's  demonstrate the validity of
 the Liberman inequality (\ref{Liberman}) first. For simplicity,  consider the case
 $\delta |_{\chi=0}=\frac{\tau-1}{2-\tau}$ of Eq. (\ref{EqXXX}), which is realized for  $\chi=0$.
 As it was mentioned  in the preceding section the QGBSTM indices  $\beta$, $\gamma'$
 and $\delta|_{\chi=0}$ for this case   coincide with the corresponding exponents of the SMM and, hence, as in the SMM case \cite{Reuter}, the Liberman inequality is fulfilled within the present model for any choice of $\alpha'$. This, however,  can be shown
 from the explicit expressions for the indices  $\beta$,  $\gamma'$ and $\delta |_{\chi=0}$, i.e. from Eqs. (\ref{beta}), (\ref{gamma}) and (\ref{EqXXX}). Using  these equations one obtains
\begin{equation}\label{EqXXXV}
\gamma'+\beta(1-\delta)=-\frac{\min\left(0,\frac{\zeta}{\varkappa}\min(\tau-1,\varkappa)-\frac{1}{\xi^T}\right)}{2-\tau}\ge0,
\end{equation}
where the  validity of the right hand side of (\ref{EqXXXV}) easily  follows now  from  the inequalities $\min(0,\ldots) \le 0$  and
$\tau<2$.
The Liberman  relation  analysis  for other values of the index $\delta$  gives the same result.
Using the  Liberman inequality  and the explicit expressions for  the QGBSTM  critical exponents  one  can  get  the following result for   the Fisher and Griffiths inequalities
\begin{equation}
\alpha'+\beta(\delta+1)\le\alpha'+2\beta+\gamma'=2+\left[\frac{\zeta}{\varkappa}-2\min\left(1,\frac{1}{\xi^T}\right)\right] \,,
\end{equation}
which holds for any value of the index $\chi$.
This equation  clearly  demonstrates that  the  Fisher and Griffiths inequalities are not obeyed  for the values of   parameters satisfying the inequality  $\frac{\zeta}{\varkappa}<2\min(1,\frac{1}{\xi^T})$.  Moreover, a  fulfillment of the Fisher scaling inequality does not guaranty that the Griffiths one is obeyed.

\begin{table}[t]
\bigskip
\begin{tabular}{|c|c|c|c|}
\hline
   \hspace*{1.0cm} & Fisher: $\alpha'+2\beta+\gamma'$ & Griffiths: $\alpha'+\beta(\delta+1)$
   & Liberman: $\gamma'+\beta(1-\delta)$ \\  \hline
   2D Ising model  &           2         &               2    &            0         \\\hline
   Simple liquids   &   $2.02\pm0.0055$   & $1.9425\pm0.0055$  &   $0.0775\pm0.0212$  \\  \hline
   3D Ising model  & $1.99996\pm0.00007$ & $2.000412\pm0.005$ & $-0.000052\pm0.002$  \\  \hline
\end{tabular}
\vspace*{0.3cm}
\caption{
Scaling relations between the critical exponents taken  from  the Table \ref{table1}.
The uncertainties  were calculated from their values given in the Table \ref{table1} using the error determination method for indirect measurements \cite{BookOnEr}.
}
%
%
\end{table}

In order to 'save' the scaling inequalities (\ref{Fisher}) and (\ref{Griffiths})  it was
suggested \cite{Fisher_70} to replace the index $\alpha'$ by $\alpha'_s$, where the index $\alpha'_s$ describes the  temperature dependence of
 the specific heat  difference
$\Delta C=(C_{\rho_H}-C_{\rho_Q})_{T=T_c}$ for two phases.
Since at  the coexistence curve  the specific heat of each phase  is defined by the  total $T$ derivative of the entropy,  then, using the  Clapeyron-Clausius equation one can  find
 $\Delta C$ as
\begin{equation}
\Delta C=\frac{T_c}{\rho_H}\frac{d}{dT}\left[T_c\frac{d\mu_c}{dT}(\rho_Q-\rho_H)\right]+
T_c\frac{\rho_Q-\rho_H}{\rho_H\rho_Q}\frac{dS_Q}{dT},
\end{equation}
where the  entropy density $S_Q$ and the baryonic densities $\rho_Q$, $\rho_H$ are calculated along the deconfinement PT line. Then  from the parameterization (\ref{xiT}) and the definition of index $\beta$  one gets
\begin{eqnarray}\label{EqXXXVIII}
\alpha'_s=
\left\{ \begin{array}{lr}
2-\beta-\frac{1}{\xi^T},  &\hspace*{0.1cm} {\rm for} \quad \frac{1}{\xi^T}<2\\
-\beta,  &\hspace*{0.1cm}{\rm for}  \quad  \frac{1}{\xi^T}\ge2\\
\end{array}\right.=\max\left(2,\frac{1}{\xi^T}\right)-\beta-\frac{1}{\xi^T}.
\end{eqnarray}
Note  that $\alpha'_s\ge0$  for $\frac{1}{\xi^T}<2-\beta$ only. Using $\alpha_s'$ from
(\ref{EqXXXVIII}) instead of $\alpha'$ in the  Fisher inequality,  one can write
\begin{equation}
\alpha'_s+2\beta+\gamma'=\max\left(2,\frac{1}{\xi^T}\right)+\left(\frac{\zeta}{\varkappa}-\beta-\frac{1}{\xi^T}\right)\ge
2-\frac{\zeta}{\varkappa}\min(0,1+\varkappa-\tau)\ge2,
\end{equation}
where the final result follows from  the obvious inequalities  $\max(2,...)\ge2$ and $\min(0,...)\le0$. Similarly, one can prove the validity of the Griffiths inequality for the index $\alpha'_s$.
Thus, the  Fisher  hypothesis \cite{Fisher_70} to replace $\alpha'$ by $\alpha_s'$ recovers the scaling  inequalities for the  critical exponents, however, it does not seem that such a suggestion  is the final solution of this problem.


%
\section{Conclusions}
\label{secconclusions}
The critical indices of the  QGBSTM with the triCEP are determined  and compared with the
critical exponents of other models.
The QGBSTM critical exponents are expressed in terms of the model parameters $\tau$,
$\varkappa$,
$\zeta$ and two newly introduced  indices $\xi^T>0$ and $\chi\ge\max(0,1-\frac{1}{\xi^T})$. The index $\xi^T$ in (\ref{xiT}) characterizes the behavior of the PT curve in  the vicinity of triCEP in the  plane of  baryonic chemical potential and  temperature, whereas
the index $\chi$ in (\ref{combinationmu}) describes the temperature dependence of a certain combination of the $\mu$-derivatives of  the model  spectrum   in the same vicinity.

Since in the FDM and SMM the index $\chi$ is implicitly set to zero, while the index $\xi^T$
either does not exists (FDM) or is fixed by other model parameters (SMM), then
the spectrum of the values of critical exponents  of the present model in more rich
compared to those models.  For  the case $\chi = 0$ and $\xi^T \le 0$ the QGBSTM reproduces the
critical exponents of the SMM with triCEP, whereas for  other choice of parameters
$\chi$ and $\xi^T$  these models belong to different  classes of universality.  Also the universality classes  of the FDM and  QGBSTM
are different since the range of values  of their  index $\tau$ is different (except for a singular case $\tau=2$): $\tau \ge 2$ in the FDM and $1 < \tau \le 2$ in the QGBSTM.
A very important result of the present work  is that, if one requires that the parameter $\varkappa$ describing the surface dependence on  the bag volume should have the values typical for  dimensions 2 and 3, then the critical indices of simple liquids and   3-dimensional Ising model  can be described only by  the SMM expressions  for  $\beta$, $\gamma'$ and $\delta$ exponents while the index $\alpha'$ differs from the SMM value.
For these sets of critical exponents a very narrow range of
 parameter $\tau = 1.826 \pm 0.02 $ is found.  Such a prediction  might be important
for experimental searches of the  QCD phase diagram endpoint since just this
exponent describes the power law  in  the volume distribution of large  bags at triCEP.

The direct calculations show that  for the standard definition of the critical index $\alpha'$
(found along  the critical isochore)  the Fisher and Griffiths scaling  inequalities are not always fulfilled, whereas the Liberman inequality is obeyed for any  values of the model parameters.
In contrast to the SMM, in which the critical isochore belongs to the boundary of the
mixed and liquid phases, the critical isochore of the present model is located
inside the mixed phase and, hence, the conditions of the Fisher theorem \cite{Fisher_64}  proving  the validity of
(\ref{Fisher}) are formally satisfied, but the  Fisher and  Griffiths   scaling inequalities are not fulfilled.
Therefore, it is quite possible that  instead of the    Fisher suggestion one should
search for
an  alternative solution of this problem and, thus,   to admit  an existence of other,
non-Fisher, universality classes  of  critical exponents
for which the right hand side of  scaling inequalities should be modified.
Hopefully, further theoretical and experimental  studies of this problem will find
its final solution.

\medskip

{\bf Acknowledgments.} The author appreciates the stimulating and fruitful  discussions with K. A. Bugaev.

\medskip



\end{document}